\documentclass[runningheads]{llncs}
\usepackage{geometry}
\usepackage[T1]{fontenc}

\usepackage{graphicx}
\usepackage{textcomp}

\usepackage{pifont}

\usepackage{url}
\usepackage{cite}

\usepackage{mathtools}
\usepackage{multirow}
\usepackage{comment}
\usepackage{booktabs}
\usepackage[ruled,vlined]{algorithm2e}
\usepackage{subfigure}
\usepackage{algpseudocode}
\usepackage{listings}

\usepackage{fancyhdr}

\usepackage{lscape}

\usepackage{cite}
\usepackage{amsmath,amssymb,amsfonts}

\usepackage{textcomp}
\usepackage{xcolor}

\def\BibTeX{{\rm B\kern-.05em{\sc i\kern-.025em b}\kern-.08em
    T\kern-.1667em\lower.7ex\hbox{E}\kern-.125emX}}

\begin{document}

\title{Markov Decision Process For Automatic Cyber Defense}

\author{Xiaofan Zhou\inst{1} \and Simon Yusuf Enoch\inst{1,2}\orcidID{0000-0002-0970-3621} \and Dong Seong Kim\inst{1}\orcidID{0000-0003-2605-187X}}
\authorrunning{Zhou et al.}

\institute{The University of Queensland, St Lucia QLD 4072, Australia\\ \and
Federal University, Kashere, Gombe State, Nigeria}

\maketitle              % typeset the header of the contribution
\begin{abstract}

It is challenging for a security analyst to detect or defend against cyber-attacks. Moreover, traditional defense deployment methods require the security analyst to manually enforce the defenses in the presence of uncertainties about the defense to deploy. As a result, it is essential to develop an automated and resilient defense deployment mechanism to thwart the new generation of attacks.
In this paper, we propose a framework based on Markov Decision Process (MDP) and Q-learning to automatically generate optimal defense solutions for networked system states. The framework consists of four phases namely; the model initialization phase, model generation phase, Q-learning phase, and the conclusion phase. The proposed model collects real network information as inputs and then builds them into structural data. We implement a Q-learning process in the model to learn the quality of a defense action in a particular state. To investigate the feasibility of the proposed model, we perform simulation experiments and the result reveals that the model can reduce the risk of network systems from cyber attacks. Furthermore, the experiment shows that the model has shown a certain level of flexibility when different parameters are used for Q-learning.

\end{abstract}
\footnotetext[1]{\textbf{Cite this article as:}\\ Zhou, X., Enoch, S. Y., and Kim, D. S. (2022). Markov Decision Process For Automatic Defense. \textit{In Information Security Applications} (WISA 2022). Lecture Notes in Computer Science, Springer, Cham.}

\section{Introduction}

Cyber-attacks have grown over the past few years to become more effective. In particular, cyber-criminals are now incorporating artificial intelligence (AI) to power cyber-attacks (e.g., deep locker \cite{stoecklin2018deeplocker}) and to outsmart conventional defense mechanisms using various approaches \cite{alavizadeh2022survey, kaloudi2020ai,park2018situational}. For instance, a group of researchers at McAfee \cite{McAfee} in their 2020 threat prediction report have predicted the potential raise of less-skilled attackers to become more powerful to create and weaponize deepfake content. In addition, they have predicted that cyber-criminals will use AI to produce convincing real data capable of bypassing many user authentication mechanisms. Besides, the current state-of-the-art defense enforcement methods require the security expert to manually deploy cyber-defenses, thus faced with uncertainties about the best countermeasures to enforce in order to achieve optimal security.

To address these challenges, we propose a novel approach to automatically select and deploy cyber defense by formulating Markov Decision Process (MDP) that reflects both attack and defense scenarios.  Specifically, we propose an automatic MDP modeling-based approach to automate defense deployment and selection using a Q-learning model (A Q-learning is a reinforcement learning policy that finds the next best action, given a current state). Here, we use the Q-learning model with the MDP framework to learn the quality of a defense action in the states. The proposed framework is divided into four phases; model initialization, model generation phase, Q-learning phase, and the conclusion phase. The model initialization phase takes a real network situation as the input and converts it into structured data; the model generation phase generates all the possible states for the MDP model using a breadth-first search algorithm;  the Q-learning phase implements a Q-learning iteration which trains the model to learn the space and update the quality for each state-action pair, and the conclusion phase searches for the optimal solutions using the Q-table trained after the previous phase. The focus of this paper is to use an AI technique to automate cyber defense and thwart attacks. The main contributions of this paper are as follows:

\begin{itemize}
    \item To design and implement an automation framework based on MDP and a deep learning algorithm for the automatic cyber defense of networked systems.
    \item To collect real network data and generate an MDP structure model based on the real data.
    \item To develop a Q-learning model which can train itself and generate an optimal defense solution.
    \item To build a testbed and to demonstrate the usability and applicability of the proposed framework based on our framework. 
\end{itemize}

The rest of the paper is organized as follows. Section \ref{sec:relatedwork} provides the related work on defense automation based on different approaches. Our proposed MDP-based framework model is  presented in Section \ref{sec:proposed_approach}. In Section \ref{sec:experiments}, we provide the experimental setup and analysis of the obtained results. We conclude the paper in Section \ref{sec:conclude}.

%Before using AI technology to deal with a real-world problem, the environment has to be modeled into a form that is readable and understandable for a computer. Markov Decision Processes (MDP) is a type of reinforcement learning in machine learning. It allows machines and software agents to "learn" about a particular problem, and to be able to output an ideal strategy to maximize the overall performance \cite{geeksforgeeks_mdp}. 

% MDP models decision making in an observable and sequential environment \cite{MDPS}. The agent changes its state in the environment, regarding the action taken by the decision-maker. Each state will have immediate and future rewards for the agent to gain, in addition to probabilities of future state transitions \cite{MDPS}. The goal of the agent is to conduct a sequence of actions that can maximize the total rewards for the problem. \\

\section{Related work}
\label{sec:relatedwork}

In this section, we briefly survey related work on defense automation for both the traditional defense and AI-based approaches.

Ray \textit{et al.} \cite{ray2005toward} proposed a framework based on UML-based use cases, state-chart diagrams, and XML to show attacker, attack actions, and the possible defense method. This work is still theoretical. Applebaum \textit{et al.} \cite{applebaum2016intelligent} developed a practical framework based on MITRE Adversarial Tactics, Techniques, and Common Knowledge (ATT\&CK)  to test for weaknesses and train defenders. In their work, they used classical planning, Markov decision processes, and Monte Carlo simulations to plan attack scenarios and to proactively move through the entire target networked systems searching for weakness and training the defenders on possible defenses to deploy.

The authors in  \cite{iqbal2020scerm} presented a framework for automating threat response based on a machine learning approach. Also, Noor \textit{et al.} \cite{noor2019machine} presented a framework for data breaches based on semantic analysis of attacker's attack patterns from a collection of threats. The focus of these papers is different from our work, as they have focused on automating threat responses from a given repository, while our proposed automation framework is based on simulation of real networks.

Zheng and Namin \cite{ZhengJianjun2018DSIN} presented a defense strategy against Distributed Denial-of-Service (DDoS) in a Software-Defined Networking (SDN) using Markov Decision Process. The authors used three parameters to model the finite set states of the MDP model, including Flow Entry Size (F), Flow Queue Size (Q), and Transmitted Packets Count (T). The rewards function is related to these three parameters F, Q, and T. Each of them has been applied with different weight factors because they have different impacts on the network. Their results show that the model can keep the flow traffic optimized and detect potential DDoS attacks at an early stage. This work also showed that the model can control how the system makes a transition by adjusting the rewards weight factor. Also, Booker and  Musman \cite{booker2020model} presented a theoretical model-based automated cyber response system, where they frame a cyber response problem as a Partially Observable Markov Decision Problem (POMDP). In another work, the authors extended their work where the POMDP is used to frame automated reasoning for defensive cyber-response that searches for a policy that maps to system states, and probabilistic beliefs.

The authors in \cite{ZhengJianjun2019MDPt} proposed a Markov Decision Process to model Moving Target Defense with the interaction between the defend and attack sides. the paper uses four states (Normal, Targeted, Exploited, Breached) with three possible defense strategies (wait, defend, reset) to describe the model. It also uses the Bellman equation and value iteration method to find out the optimal policy for each state. Their result demonstrated how much impact the cost will have on the optimal policy and how that will help the defender to make better defense strategies. Other authors such as \cite{enoch2022practical,enoch2022integrated} developed a blue team framework that can perform cyber defense generation, defense enforcement, and security evaluation using a defined workflow. However, the work did not use any AI technique to enhance system attack learning or to thwart cyber attacks.

\section{The Proposed Approach}
\label{sec:proposed_approach}

In this section, we describe the proposed framework for automatic cyber defense based on MDP. The workflow of the framework comprises of four phases; Initialization Phase, Generation Phase, Q-learning Phase, and Conclusion Phase. We explain them in detail as follows.

\subsection{Model Initialization Phase}

The first phase is the initialization phase. During this phase, the program takes some real network situations as the inputs. These inputs need to be recognized and transformed into programmed data and later implemented into the MDP model. Here, the more detailed the description of the network situation is, the more complex the model will become. 

% ****************** Section ******************
\subsection{Model Generation Phase}

The second phase is the model generation phase. During this phase, the program will generate all the possible states for the MDP using the input data collection from the previous phase. To guarantee all the states will be visited in a well-designed order, it is necessary to have a traversal method (and the Breadth-First Search (BFS) algorithm will be used in this phase). Algorithm \ref{alg:initial} and Algorithm \ref{alg:generatenextstate} are used for the model generation, including the generation of the next state and the defense states.

\begin{algorithm}[ht]\scriptsize
\SetAlgoLined
 queue.add(initialState)\;
 \While{queue not empty}{
  currentState $\leftarrow$ queue.pop()\;
  states.add(currentState)\;
  GenerateNextState(currentState)\;
 }
 \caption{Initialize States}
 \label{alg:initial}
\end{algorithm}

\begin{algorithm}[ht] \scriptsize
\SetAlgoLined
 \tcc{Generate Attack States}
 \If{attackPath is None}
  {
    \For{host $\leftarrow$ adjacentHost}
      {
        \If{!host.compromised \& host.hasVulnerabilities}
        {
           state $\leftarrow$ AttackAction(host)\;
           queue.add(state)\;
        }
      }
  }
 \Else
  {
     host $\leftarrow$ GetNextHostOnPath()\;
     \If{!host.compromised \& host.hasVulnerabilities}
    {
       state $\leftarrow$ AttackAction(host)\;
        queue.add(state)\;
    }
  }
  \tcc{Generate Defense States}
  \For{action $\leftarrow$ defenseActionsList}
  {
     state $\leftarrow$ DefenseAction(action)\;
    queue.add(state)\;
  }
 \caption{Generate Next State}
 \label{alg:generatenextstate}
\end{algorithm}

There are two major assumptions during the model generation phase. Firstly, the attacker can only attack the host which is next to a compromised host or public internet. For example, if the attacker attempted to compromise one host in the network, this is only possible to happen when there is at least one neighbor host compromised, or the host is directly connected to the public internet. Secondly, there is no value to patch vulnerabilities on a host that has already been compromised. Once a host is marked as compromised in the model, it is assumed that the data on the host has already been fully breached or the host has already been controlled.

After all the states have been generated, a transition table will be constructed. The table has size \textit{s} by \textit{s} where s is the number of all states. Each cell contains transition information between the row state and the column state, or none represents no transition available between two states. The transition information includes data such as action, success rate, reward after success transition, and reward after the fail transition.

% ****************** Section ******************
\subsection{Q-learning Phase}

The third phase is the Q-learning phase. During this phase, the model will keep learning the space until the iteration is over. Before starting the learning process, a Q-value table will need to be initialized with rows and possible actions, and columns as all generated states. %Unlike the figure, the q values in the cells are all zero initially. 
Here, each q-value represents the "quality" of a state and action pair. During this learning phase, the Q-table will keep updating until it has reached the maximum iteration. %The general flow for the Q-learning is shown in Figure \ref{fig:Qlearning}.

\begin{comment}
\begin{figure}[ht]
	\centering
	\includegraphics[width=0.5\linewidth]{fig/QLearning.png}
	\caption{Q-learning Flow}
	\label{fig:Qlearning}
\end{figure}
\end{comment}

Four parameters are needed for the Q-learning; learning rate, epsilon, epochs, and gamma ($\gamma$) or discounted factor which is ranging from 0 to 1. The $\gamma$ parameter decides how important the future rewards will be. It is also used to approximate the noise in future rewards. The Q-learning phase is described by Algorithm \ref{QlearningTrain}.  In this phase, if gamma is close to one, it means the agent mostly considers the future rewards while being willing to delay the immediate rewards. If gamma is close to zero, it means the agent will mostly only consider the immediate rewards.

\begin{algorithm}[ht] \scriptsize
\textbf{Input:} gamma, lrnRate, epsilon, maxEpochs\;
\For{i in range(maxEpochs)}
{
    currS $\leftarrow$ 0\;
    \While{True}
    {
     \tcc{Decide to explor or exploit}
        \If{random.uniform(0, 1) < epsilon}
        {
            action $\leftarrow$ GetRandomNextAction(currS)\;
        }
        \Else
        {
            action $\leftarrow$ GetMaxAction(currS)\;
        }
        
        nextS = GetStateFromAction(action)\;
        \tcc{Finish if no following state}
        \If{nextS is None}
        {
            break\;
        }
        
        \tcc{Whether the action is successful or fail}
        \If{random.uniform(0, 1) < trans[currS][nextS].rate}
        {
            reward $\leftarrow$ rewards[currS][nextS].success\;
        }
        \Else
        {
            reward $\leftarrow$ rewards[currS][nextS].fail\;
            nextS $\leftarrow$ currentS\;
        }
        
        nextA $\leftarrow$ GetMaxNextAction(nextS)\;
        futureQ $\leftarrow$  QTable[nextS][nextA]\;
        
        \tcc{Update Q Table}
        QTable[currS][action] $\leftarrow$ QValueCalculation()\;
        
        currS $\leftarrow$ nextS\;
    }
}
 \caption{QLearningTrain}
 \label{QlearningTrain}
\end{algorithm}

Equation \eqref{q_value} shows the detail calculation for the function \textit{QValueCalculation()}.

\begin{equation} \label{q_value}
Q(s, a) = ((1 - \alpha) * Q(s, a)) + (\alpha * (reward + (\gamma * Q(s', a')))
\end{equation}

Here, the Q-learning needs to make sure every q-value has been updated with sufficient times to reflect the actual quality. The agent can increase the number of iterations (epochs) to increase the overall updated times. The agent can also adjust the epsilon to balance between exploration and exploitation. %An analysis of those parameters will be further discussed in the Result Discussion section.

\subsection{Conclusion Phase}

After the Q-learning process has been completed and the Q-table has finished its updates, the process will enter the conclusion phase. The main task in this phase is to find the optimal solution(s) for the current network state. 

\section{Experimental Setup}
\label{sec:experiments}

In this section, we use a real network to illustrate the framework used for the attack and defense scenarios.

\begin{figure}[ht]
	\centering
	\includegraphics[width=0.4\linewidth]{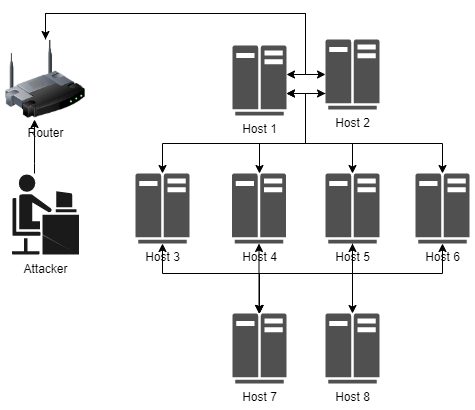}
	\caption{Real Network Structure}
	\label{fig:RealNetwork}
\end{figure}

\textbf{The network and attack model}: The network structure is shown in Figure \ref{fig:RealNetwork}. The network consists of 8 hosts, named host 1 - host 8. 
The network has a router that controls access between the networked hosts. Hosts in the network have vulnerabilities that may or may not be patchable. Table \ref{tbl:vul_info} shows the vulnerabilities of each host.  In the Table, we use $V_i$ to denote vulnerability ID, CVSS Score for Common Vulnerability and Scoring System Base Score, and Patch cost for the cost of patching vulnerabilities. The CVSS score is based on the severity scores provided by National Vulnerability Database \cite{CVSS}, and we assume the patch cost value.
We assume an attacker is located outside the network. The attacker is trying to compromise the host in the internal network. The attacker can directly connect to host 1 and host 2. 

In our model, we represent the connections between hosts with links. For example, the hosts ($h_i$) information is going to be recorded as a list such as [$h_1$, $h_2$, ...,$h_n$], and links will be represented as [($h_1$, $h_2$), ($h_2$, $h_1$), ($h_1$, $h_3$) \dots]. In a real situation, the network connections between two hosts are not always bi-directional. It is possible for a host to stop receiving packages from another host while it is still able to send packages to that host. Therefore, all the links recorded in the program are uni-directional.

\begin{table}[ht] \scriptsize
\centering
\label{tbl:vul_info}
\caption{Hosts and Vulnerabilities Information}
\begin{tabular}{|c|c|c|c|} 
     \hline
     Host Address & Vulnerability ID & CVSS Score & Patch Cost \\
     \hline
     172.16.0.1 & $V_1$ & 4.3 & 8.0 \\ 
     172.16.0.2 & $V_2$ & 2.1 & 5.0 \\
     172.16.0.3 & $V_3$ & 10.0 & 6.5 \\
     172.16.0.4 & $V_4$ & 4.3 & 3.5 \\
     172.16.0.5 & $V_5$ & 7.5 & 4.5 \\
     172.16.0.6 & $V_6$ & 8.8 & 5.0 \\
     172.16.0.7 & $V_7$ & 8.8 & 6.0 \\
     172.16.0.8 & $V_8$ & 6.1 & 7.0 \\
     \hline
\end{tabular}
\end{table}

\textbf{Defense model}:
Since it is infeasible to patch all vulnerabilities in real network environments, we assume only a few defense options can be selected for possible defense. We explain each of the defenses as follows and we show the available defense strategies in Table \ref{tbl:defense}.

\begin{itemize}
    \item BLOCK(target, sub-target): Block port action takes two parameters, target, and sub-target. Target tells the model to block port on which host, while sub-target indicates which host should be blocked connection from. For example, command BLOCK(172.16.0.2, 172.16.0.1) represents the host 172.16.0.1 should block port from host 172.16.0.2.
    \item PATCH(target, vulnerability): Patch action takes two parameters, target, and vulnerability. For example, command PATCH(172.16.0.3, V3) represents patching vulnerability V3 on host 172.16.0.3.
\end{itemize}

\begin{table}[ht] \scriptsize
\centering
\caption{Available Defenses Options}
\label{tbl:defense}
\begin{tabular}{|c|l|} 
     \hline
     Defenses ID & Defense Detail \\
     \hline
     D1 & Block port to Router on 172.16.0.1 \\ 
     D2 & Patch V7 on 172.16.0.7 \\
     D3 & Block port to Router on 172.16.0.2 \\
     D4 & Block port to 172.16.0.7 on 172.16.0.3 \\
     D5 & Patch V4 on 172.16.0.4 \\
     D6 & Patch V6 on 172.16.0.6 \\
     \hline
\end{tabular}
\end{table}

\subsection{Results and analysis}
In this section, we use the network scenario described to illustrate the phases of the framework with their results.

\subsubsection{Initialization phase}
One of the features and an MDP-based model assumes that the environment is fully observable and known by the agent. In this phase, the hosts and vulnerabilities are collected and provided as input to the model, it is presumed that the data collected have covered all the hosts and vulnerabilities in the space. For this experiment, the following network data were collected:

\begin{itemize}
    \item Host Address: The IP address for the host. This data is treated as the identifier for each host in the model.
    \item CVSS Score: This value is collected from NVD. The number has a range from 0 to 10. The higher the number, the more severe the vulnerability is when it is compromised by attackers. This number will be used as a negative offset in the model's rewards calculation for state transition, particularly for an "attack" transition.
    \item Vulnerability ID: is an identifier for each vulnerability. Hosts can have more than one vulnerability.
    \item Patch Cost: is a number that represents the total cost of patching the vulnerability on the host. The number has a range from 0 to 10. For example, the cost of patching $V_1$ is 8.0 and the cost of patching $V_3$ is 6.5. The number will be used as a negative offset in the model's rewards calculation for state transition, particularly for a "patch" transition.
\end{itemize}

Attack Path is an optional input for the model, it decides whether the model is trying to solve a more particular problem or a wide-ranged problem. If the attack path is given, the attacker will only attack the host which is on the path. If the attack path is not specified in the model, the model will assume the attacker will attack any feasible host for the attacker to attack. The attack patch is an important element during the model generation phase. For this experiment, the following attack path (Figure \ref{fig:AttackPath}) is used:

\begin{figure}[ht]
	\centering
	\includegraphics[width=0.5\linewidth]{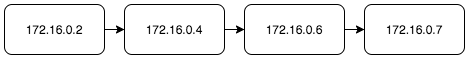}
	\caption{The attack path}
	\label{fig:AttackPath}
\end{figure}

\subsubsection{Model Generation Phase}

\begin{figure*}[ht]
	\centering
	\includegraphics[width=0.8\linewidth]{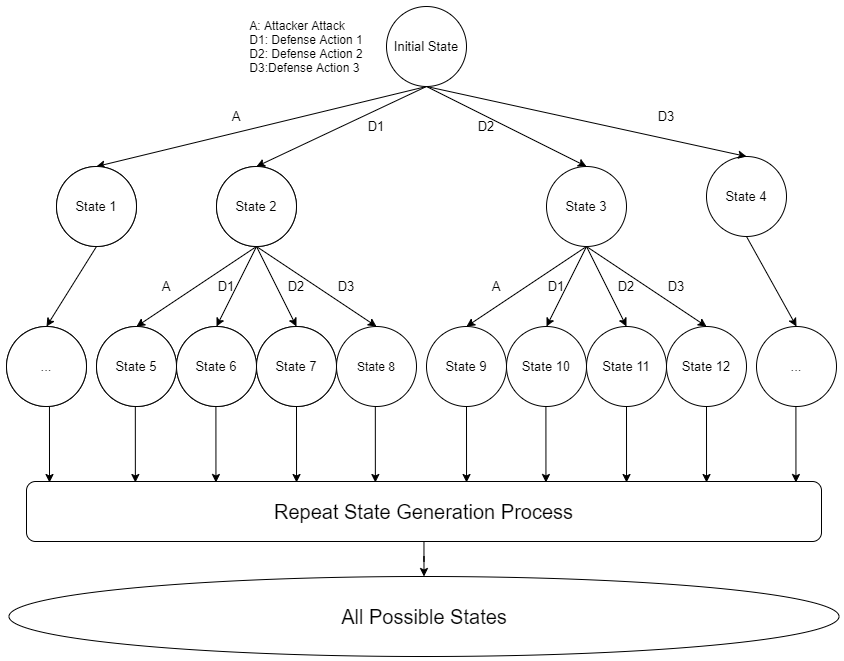}
	\caption{State Generation Diagram}
	\label{fig:StateGeneration}
\end{figure*}

Figure \ref{fig:StateGeneration} shows how the BFS algorithm starts exploring the space from its root node, which corresponds to the initial state in the model. It explores the next level of states using all possible attack and defense actions. There are four possible actions to perform in the initial state in the figure, and thus it expands its branches to those four states. The algorithm will finish exploring all of the neighbor states at the same level before moving to the next depth level.

Each node in the BFS exploration tree will be visited only once, however it is still possible for two nodes to have the same state. This is because doing different actions in a different sequence is possible to result in the same state. Therefore, it is necessary to check duplications before adding the node to the state set in the MDP model. 

Here, the attack path is an important element during the exploration of the BFS algorithm. If the attack path is not specified, the algorithm assumes the attacker will attack any feasible host (i.e., the host being attacked is adjacent to a compromised host and the host has at least one vulnerability.) Not specifying the attack path will add complexity and run time for the model generation phase.

\subsubsection{Q-learning Phase}

In this phase, the Q-Value Table shown in Figure \ref{fig:QTable} is initialized. Initially, the Q values in the cells are all zero. Each Q-value represents the "quality" of a state and action pair.

The parameters for Q-learning iteration is listed as follows:
\begin{itemize}
    \item $\gamma$ (Discount Factor) : 0.9
    \item $\alpha$ (Learning Rate) : 0.1
    \item $\epsilon$ : 0.7
    \item epochs : 5000
\end{itemize}

For this simulation experiment, 1492 possible states have been generated in total. After the model finishes its process, all the output data generated were written into a text file, including the optimal defense solutions and the full q-table after training. The q-table is recorded into n lines where n is the number of possible states. Each line represents the data for each state.
One part of the line is used to describe the situation of the state, (1) "Compromised Hosts" gives a list of hosts that have been compromised by the attacker; (2) "Links" provide a list of existing connections between the hosts, blocked links will not be included; (3) "Vulnerabilities" give a list of existing vulnerabilities, patched vulnerabilities will not be included. Another part is the q-value for each action at this state. For example, the q-value for action 4 at this state is -2.1.

\begin{quote} 
          Compromised Hosts: [0, 2, 6]\\
          Links: [(0, 1), (0, 2),(1, 2), ..., (7, 5), (7, 6)]\\
          Vulnerabilities: [V1, V3, V4, V5, V6, V7]\\
          Q-Values: -9.298, -6.19, -7.89, -2.1, -10.69, -5.39, -6.89
\end{quote}

If an attack path (Figure \ref{fig:AttackPath}) is provided to the model, the optimal defense sequence for this network is D3 (Block port to 172.16.0.0 on 172.16.0.2). From the network structure perspective, we can see that if host2 blocks connection from host0, then according to the pre-defined attack path, the attacker will not be able to make any attack action. Therefore, the network is secured after performing only one defense action- D3. 

 From the q-table perspective (Table \ref{tbl:q-table-path}), D3 has the largest q-value at the State0 (Initial State), so D3 is added to the output sequence. For State5 (the state after performing D3 at State0), the q-value for attack action is 0 which is larger than any other defense action. Therefore, the model concludes that there is no need to perform any defense actions, and the search ends.

\begin{table*}[ht] \scriptsize
\centering
\caption{Q-Table for Real Network Example with Attack Path (Partial)}
\label{tbl:q-table-path}
\begin{tabular}{|p{3cm}|c c c c c c c|} 
     \hline
     State & Attack Action & D1 & D2 & D3 & D4 & D5 & D6\\
     \hline
     State0 (Initial State) & -9.298 & -6.19 & -7.89 & -2.1 & -10.69 & -5.39 & -6.89\\ 
     \hline
     State5 (T(State0, D3) = State4) & 0.0 & -4.3 & -6.0 & 0.0 & -8.8 & -3.5 & -5.0\\
     \hline
\end{tabular}
\end{table*}

If the attack path is not provided to the model, the optimal defense sequence for this network is D3-D1. From the network structure perspective, if host1 and host2 both block connection from host0, then the rest of the network is fully protected because host1 and host2 are the only passes where the attack can proceed its attack. From the q-table perspective (Table \ref{tbl:q-table-withoutpath}), D3 has the largest q-value at the State0 (Initial State), so D3 is added to the output sequence. D1 has the largest q-value at State5 (Initial State), so D1 is then added to the output sequence. For State29 (the state after performing D1 at State5), the q-value for attack action is 0 which is larger than any other defense action.

\begin{table*}[ht] \scriptsize
\centering
\caption{Q-Table for Real Network Example without Attack Path (Partial)}
\label{tbl:q-table-withoutpath}
\begin{tabular}{|p{3cm}|c c c c c c c|} 
     \hline
     State & Attack Action & D1 & D2 & D3 & D4 & D5 & D6\\
     \hline
     State0 (Initial State) & -9.298 & -6.19 & -9.296 & -5.97 & -12.363 & -7.283 & -8.274\\ 
     \hline
     State4 (T(State0, D3) = State5) & -5.33 & -4.3 & -9.411 & None & -12.415 & -7.203 & -8.667\\
     \hline
     State29 (T(State5, D1) = State29) & 0.0 & 0.0 & -5.948 & 0.0 & -8.622 & -3.478 & -4.946\\
     \hline
\end{tabular}
\end{table*}

The output result may significantly depend on the network situation, such as the cost of patching a vulnerability, the cost of blocking ports on a host, or the damage to a host after being attacked. For some network systems,  blocking ports on host1 (D1) may result in further damage to the organization's service, because it not only blocks the attacker but also blocks all other normal users from accessing. In that case, the cost of D1 will be raised significantly, and as a result, the optimal defense sequence may not include D1. The model may choose other alternative defense strategies, such as patching vulnerabilities on host4 (D5), to minimize the damage.

\subsubsection{Conclusion Phase}

In this phase, all the q-values in the table are negative since the implementation of cyberdefense is a costly task. Either patching a vulnerability or host compromised puts a negative effect on the whole system. It is impossible to profit and earn positive rewards. 

\begin{figure}[ht]
	\centering
	\includegraphics[width=0.8\linewidth]{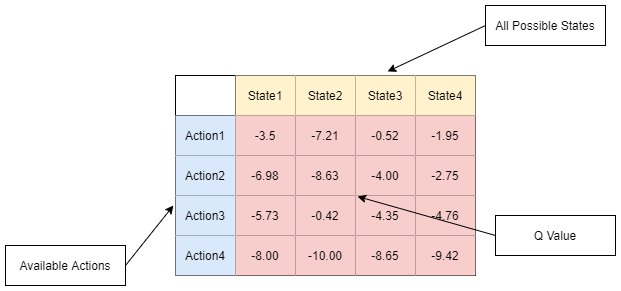}
	\caption{Q-Value Table}
	\label{fig:QTable}
\end{figure}

Finding the optimal strategy for one certain state can be achieved by looking at the corresponding q-value in the q-table. The larger the q-value is, the less damage the action will result. For example in Figure \ref{fig:QTable}, Action1 has the largest q-value in the State1 column, which means Action1 theoretically is the best action to take at State1. On the contrary, Action 4 is the worst action to take.

Finally, the model will look for a sequence of actions from the initial state. The model will keep searching through the q-table by using the following steps:
\begin{enumerate}
    \item Find the best action in the initial state, add the action to the sequence;
    \item Go to the consequence state with the action. For example, performing Action1 in State1 will result in State2.
    \item Find the best action at that state and add the action to the sequence. Keep iterating step 2 \& 3 until there is no following states, or the q-value shows there is no need to do any defense actions. (When the q-value of action attack is larger than any other defense actions, it means that performing any defense actions will be redundant and cause more damage to the system. That is when there is no need to do any defense actions).
\end{enumerate}

After performing the above steps, the model will output a sequence of actions such as Action1-Action3-Action2.
This action sequence is the solution that optimized the rewards, therefore minimizing the overall cost and damage to the network system. The initial state can be replaced by any state for this searching mechanism, which allows the agent to find an optimal solution in any situation in the network.

\subsection{Effect of Q-learning Parameters on Optimal Reward}

Different parameters' values may have a significant effect on the output result of the model. In this section, Q-learning parameters such as discount factor, epsilon, and iterations are investigated. This section will assess their performance with different data, and a suitable combination of parameters should be concluded to maximize the overall performance of the model.

\begin{figure}[ht]
	\centering
	\includegraphics[width=0.6\linewidth]{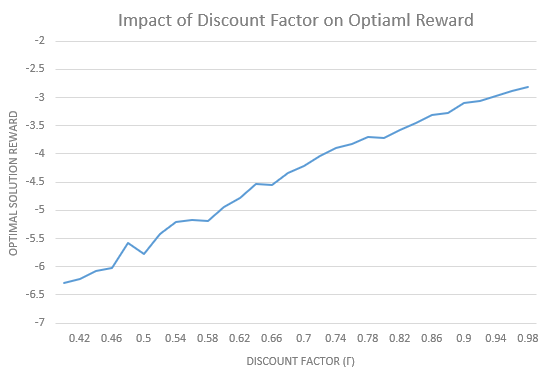}
	\caption{Impact of Discount Factor on Optimal Reward}
	\label{fig:DiscountFactor}
\end{figure}

Discount Factor ($\gamma$) in Q-learning, $\gamma$ $\in$ (0, 1), indicates the importance of the future rewards compared to the immediate rewards. If $\gamma$ is larger, it means the agent considers the future rewards more and is willing to delay the immediate rewards. As figure 4.3 shows, the optimal solution reward grows almost linearly as the discount factor increase. As the model digs deeper into space, the more certain it realizes that the optimal solution has a better effect on defending the network system. This explains why the damage becomes less when the discount factor increases.

Secondly, epsilon ($\epsilon$) is a factor that balances exploration and exploitation. If $\epsilon$ is larger, the agent will have more possibility to explore the space (i.e. to choose action randomly). If $\epsilon$ is lesser, the agent will be more likely to choose the action with the highest q-value. Changing $\epsilon$ has a minor effect on the overall optimal rewards value, therefore in this experiment, the percentage improvement of the optimal solution's reward, compared to the rewards of not defending, is used as an index to test the performance of the model. The percentage is calculated by equation \eqref{improv_per} (where OSR is Optimal Solution Reward, NDRs is No Defend Rewards).

\begin{equation}\label{improv_per}
\textit{Improvement Percentage} = - \frac{OSR - NDRs}{NDRs}
\end{equation}

\begin{figure}[ht]
	\centering
	\includegraphics[width=0.6\linewidth]{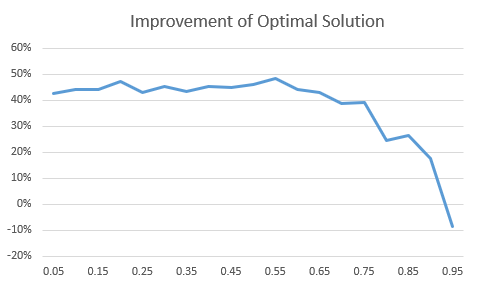}
	\caption{Improvement of Optimal Solution with Different Epsilon}
	\label{fig:Eplison}
\end{figure}

Since the rewards are all negative, so the result needs to be negated. Figure \ref{fig:Eplison} shows the improvement almost stays at 40\% to 50\% level when $\epsilon$ is less than 0.75. However, after the $\epsilon$ gets larger than 0.75, the improvement drops substantially and even decreases below 0\% (the optimal reward is less than the attack reward) at 0.95. This shows when $\epsilon$ gets larger than a certain point, the agent tends to explore more paths. As a result, the agent did not give much weight to the optimal solution, therefore, reducing the difference between every action.

\begin{figure}[ht]
	\centering
	\includegraphics[width=0.7\linewidth]{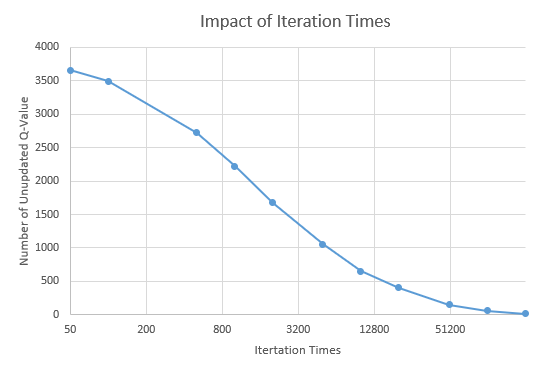}
	\caption{Improvement of Iteration Times on the Number of Un-Updated Q-Value}
	\label{fig:Iteration}
\end{figure}

 Lastly, iteration times (epochs) represent the maximum number of iterations the Q-learning will iterate. As the iteration times increase, the overall quality and completeness of the output result are also increased. However, larger iteration times can also increase the run time for the model. Therefore, it is necessary to find a suitable number of iteration times that allows both a decent run time and an acceptable quality of the output results.

As figure \ref{fig:Iteration} shows that the iteration times increase, the number of un-updated q-value decreases (Note that the x-axis is in log-scale). The un-updated q-value means the q-value in the q-table that has not been updated once. Since the q-table for the experiment has a size of  13594 q-values, likely, some of them are not updated during the process. The pattern of the graph is similar to a logarithm equation. When epochs are relatively small, the un-updated q-values decrease substantially. When epochs are relatively large, the un-updated q-values only decrease a small amount. That is because when there is more and more q-value being updated, the probability for the agent to reach an un-updated q-value becomes less.

Although the parameters for Q-learning can vary between different tasks, an appropriate range of those parameters has been concluded for a network defense problem. 

\begin{itemize}
    \item Discount Factor ($\gamma$): The experiment reveals that the larger the $\gamma$ is, the more rewards will be received from the optimal solution. However, it is also not proper to weigh too much on the future rewards, since the immediate rewards still need to be considered in some cases. In summary, the experiment suggests a range of 0.8 to 0.9 for the discount factor.
    \item Epsilon ($\epsilon$): The experiment shows that if $\epsilon$ increases over 0.75, the overall difference between the optimal solutions and other solutions will be reduced. Therefore, it will become hard to distinguish between a "good" and "bad" action. While if $\epsilon$ gets too small, the agent will be less likely to find alternative strategies that can further reduce the overall damage to the network system. As a result, the experiment suggests a range of 0.5 to 0.7 for the $\epsilon$.
    \item Iteration Times (epochs): The experiments prove that as epochs increase, the overall quality and completeness of the output result is also increased. However, the efficiency of improving the result decreases, and the run time raises when epochs increase to a larger number. Therefore, the experiments suggest a range of 5000 to 10000 for the epochs.
\end{itemize}

\subsection{Model Efficiency Experiments}

In the area of cyber defense, algorithm efficiency is also a key element to determine whether the system can successfully defend against the attacker. If the attacker's efficiency is better than the defense side, the optimal solution output from the model may be no longer applicable to the environment. 

The experiment uses a network with different numbers of hosts to test the efficiency of the model. Normally, if there are more hosts in the network, the model will become more complex and there will be more possible states to generate. As the state number increases, the complexity of training the model also increased.

\begin{figure}[ht]
	\centering
	\subfigure[]{
		\label{fig:Eff_generate}
		\includegraphics[width=0.47\textwidth]{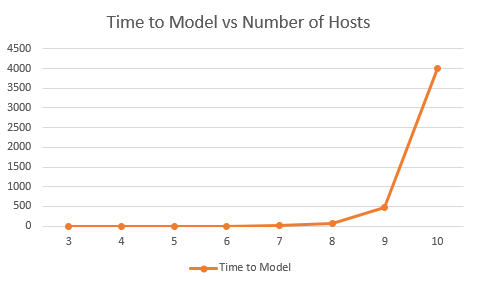}
	}
	\subfigure[]{
		\label{fig:Eff_Train}
		\includegraphics[width=0.47\textwidth]{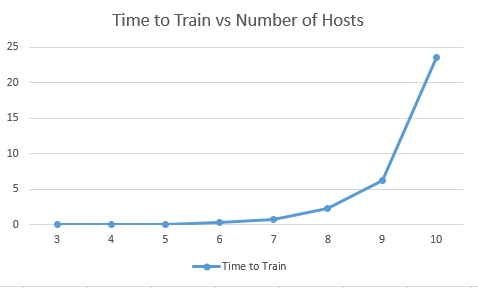} 
	}
	\caption{Time Efficiency for Model Generation }
	\label{fig:Efficiency}
\end{figure}

Figure \ref{fig:Efficiency} shows when the number of hosts is less than 8, the time to generate and the time to train the model does not increase much as the number of hosts increases. The time for model generation stays under 1 minute and the time for training stays under 5 seconds. However, when the number of hosts is greater than 8, the time spent starts to grow exponentially. On the other hand, the time to train the model is relatively faster than the time to generate the model. 

\section{Conclusion and Future Work}
\label{sec:conclude}

In this paper, we have presented an MDP-based optimal solution model for cyber defense. The model is composed of four sequential phases. The model initialization phase takes some real network situation as the input and converts it into structured data; the model generation phase generates all the possible states for the MDP model using a breadth-first search algorithm; the Q-learning phase implements a Q-learning iteration which trains the model to learn the space and update the quality for each state-action pair; the conclusion phase searches for the optimal solutions using the q-table trained after the previous phase.
Real network simulation experiments have been done to test the usability and functions of the model. The result demonstrates the model can reduce the attack impact on the network system from a cyber-attack, in either network structure prospective or q-table perspective. In the future, we plan to add more defense actions to the model. Another potential development for the model is to make it a POMDP (Partially observable Markov decision process). Besides, We plan to collect more usable and real data from a bigger network.

\bibliographystyle{splncs04}
\bibliography{mybibfile}

\end{document}